\documentclass[twocolumn,english]{revtex4-1}
\usepackage[T1]{fontenc}
\usepackage[latin9]{inputenc}
\usepackage{amsmath}
\usepackage{amssymb}
\usepackage{graphicx}
\usepackage{wasysym}
\usepackage{esint}

\makeatletter
\usepackage{soul}
\usepackage{color}
\usepackage{babel}

\makeatother

\usepackage{babel}
\begin{document}
\title{Universal Features of Annealing and Aging in Compaction of Granular
Piles }
\author{Paula A. Gago$^{1}$ and Stefan Boettcher$^{2}$}
\affiliation{$^{1}$Department of Earth Science and Engineering, Imperial College,
London SW7 2BP, United Kingdom~~~\\
 $^{2}$Department of Physics, Emory University, Atlanta, GA 30322,
USA}
\begin{abstract}
We explore the compaction dynamics of a granular pile after a hard
quench from a liquid into the glassy regime. First, we establish that
the otherwise athermal granular pile during tapping exhibits annealing
behavior comparable to glassy polymer or colloidal systems. Like those
other systems, the pile undergoes a glass transition and ``freezes''
into different non-equilibrium glassy states at low agitation for
different annealing speeds, starting from the same initial equilibrium
state at high agitation. Then, we quench the system instantaneously
from the highly-agitated state to below the glass transition regime
to study the ensuing aging dynamics. In this classical aging protocol,
the density increases (i.e., the potential energy of the pile decreases)
logarithmically over several decades in time. Instead of system-wide,
thermodynamic measures, here we identify the intermittent, irreversible
events (``quakes'') that actually drive the glassy relaxation process.
We find that the event rate decelerates hyperbolically, which explains
the observed increase in density when the integrated contribution
to the downward displacements is evaluated. We argue that such a hyperbolically
decelerating event rate is consistent with a log-Poisson process,
also found as a universal feature of aging in many thermal glasses. 
\end{abstract}
\maketitle
In thermodynamic systems, physical aging arises when disordered systems
fall out of equilibrium after a sudden quench through the glass transition
into a glassy regime where relaxation time-scales begin to exceed
experimental time-scales at low temperature, high density, low shear,
etc~\cite{Liu98}. The experimental observation of ``anomalous''
non-equilibrium events~\cite{Buisson03,Buisson03a,Bissig03,Yunker09,Kajiya13,Zargar13,Tanaka17}
have established intermittency as key property of the relaxation dynamics
commonly referred to as ``aging''~\cite{Struik78}, in a wide variety
of glassy materials. Spin-glass thermo-remanent magnetic data~\cite{Sibani06a},
magnetic flux creep in type-II high-$T_{c}$ superconductors~\cite{Oliveira05},
and particle motion data in dense colloids~\cite{BoSi09,Robe16} have
all been interpreted using the statistics of ``quakes'', i.e., rare,
localized events which lead the system from one of its metastable
states to the next.

The classic experiments on the ``Chicago pile''~\cite{Nowak98,Nowak97}
have established the slow dynamic nature of the relaxation in tapped
granular piles. Unlike for fluctuating thermodynamic systems~\cite{Nowak97},
a tap is needed for a cycle of momentary acceleration of the grains, followed by
complete dissipation. By measuring the density evolution of a
tapped column of grains, Ref.~\cite{Nowak98} reported that, for
low intensities taps, changes in density are logarithmic in time (measured
in taps). Remarkably, such logarithmic behavior is also characteristic
of relaxation in thermal systems~\cite{Hutchinson1995,Amir2012,roth2016polymer}.
For the pile, this logarithmic density evolution has been successfully
modeled using the parking-lot model for compaction~\cite{Nowak98,BenNaim98},
which has been noted as an example of event-driven dynamics~\cite{Sibani16}.
The events driving the long-time relaxation behavior in granular experiments
are unclear. Such granular piles have been shown to fall out of equilibrium
when the energy infused by tapping is insufficient to break all contacts
between grains during each tap~\cite{gago2017ergodic}. How changes
within the pile, e.g., in its contact network, relax the free energy
and increases its density remains largely unexplored.

\begin{figure}
\centering \includegraphics[width=1\columnwidth]{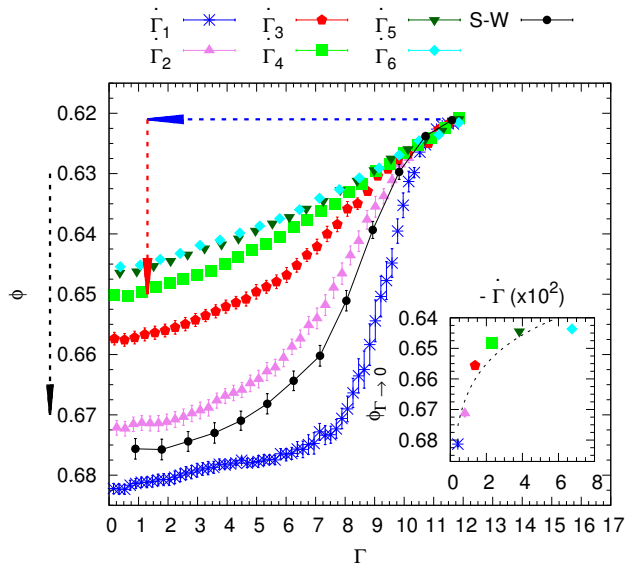}
\caption{Average packing fraction $\phi$ as a function
of the dimensionless acceleration $\Gamma$ during different annealing
protocols, all initiated after sufficient equilibration at
$\Gamma_{{\rm high}}=12$. For reference, a protocol (S-W) with step-wise 
decrements of $-\Delta\Gamma=0.92$, imposed after $150$
taps at each $\Gamma$, similar to the Chicago experiments~\cite{Nowak97,Nowak98},
was adopted to create the ``reversible branch''. In
the other simulations, the acceleration $\Gamma$ was \emph{gradually} reduced
after each tap at various constant decrements such that $-\dot{\Gamma}_{i}=v_{i}\times10^{-2}$
with $v_{1}=0.279 $, $v_{2}=0.808 $, $v_{3}=1.373 $,
$v_{4}=2.334 $, and $v_{5}=3.83$,
$v_{6}=6.747$. Results
are averaged over $10$ independent realizations, with error bars
corresponding to the standard error of the mean. 
In turn, the arrows indicate the aging protocol: an instantaneous
quench from $\Gamma_{{\rm high}}$ to a finite $\Gamma_{q}=1.3$ (arrow
left), well below the glass transition, is executed after which system properties are traced as a function of time/taps (arrow down), as shown in Figs.\ \ref{fig:agingdensity}-\ref{fig:changingneighbors}.
Inset: Plot of the terminal plateau value $\left\langle \phi{}_{\Gamma\to0}\right\rangle $
reached for $\Gamma<1$ in main panel as a function of annealing speed
$-\dot{\Gamma}.$ The dotted line is $\left\langle \phi{}_{\Gamma\to0}\right\rangle \sim0.696-0.026(-\dot{\Gamma})^{1/4}$,
merely to guide the eye. }
\label{fig:Gammadot} 
\end{figure}

In this Letter, we study the relaxation of the Chicago pile in molecular
dynamics (MD) simulations and show that the ensuing non-equilibrium
aging phenomenology is surprisingly similar to other glassy systems.
Density fluctuations are induced athermally through a cycle of taps
imposed externally to briefly accelerate the grains upward. Kinetic
energy is completely dissipated through friction before the application
of a new tap. First, we establish that the granular pile exhibits
annealing behavior comparable to glassy polymer or colloidal systems,
shown in Fig.\ \ref{fig:Gammadot}. To that end, we anneal the pile
repeatedly from an equilibrated, highly agitated regime into a completely
jammed state over a wide range of fixed speeds by which the acceleration amplitude
is gradually reduced. The system falls apparently out of equilibrium below a
certain level of agitation, akin to a glass transition~\cite{Hutchinson1995,Debenedetti01,Zhu13,roth2016polymer,Gedde19},
by attaining densities that systematically vary with the speed. Then,
we study the aging dynamics following the conventional protocol of
quenching instantly, at infinite speed, from the equilibrated into
the nearly jammed regime of agitation, well below the observed glass
transition but still sufficiently agitated to discern its relaxation.
The evolution of density on average reproduces the logarithmic rise
found in the Chicago experiment. We can attribute this slow rise to
a hyperbolic deceleration in the rate of irreversible events by which
grains progressively solidify in their neighborhood matrix. These
events are signaled by a geometric rearrangement of a grain reaching
a new record in its number of contacts.

We employ a numerically implementation inspired by the experimental setup
used for the Chicago experiment. Our pile consists of $60\,000$ bi-disperse
spheres, {[}$1$-$1.02${]}mm diameter, contained in a vertical cylinder
of $2.4$cm diameter. The resulting column height is about $11.5$cm.
A minute bi-dispersity, in equal proportion, was introduced in order
to prevent crystallization and the grain diameter ratio was chosen
to avoid segregation.

We employ molecular dynamics (MD), in particular, the implementation of the soft spheres~\cite{brilliantov1996model} model  provided by the $LIGGGTHS$~\cite{goniva2012influence} open source software. Within $LIGGGHTS$,
we employ a $Hertz$ model for the grain-grain and grain-wall contact forces (with Young
modulus $Y=6\times10^{6}$Nm$^{-2}$, Poisson ratio $n=0.3$, restitution
coefficient $\epsilon=0.5$ and friction coefficient $\mu=0.5$). 
Particle densities equal $2500$Kg/m$^{3}$.
% The selection of these parameters ...

Taps consist of moving the whole container along a semi-sinusoidal
wave pulse of amplitude $A$ and angular frequency $\omega$. 
A new tap will be applied to the system
only after the system has attained a mechanically stable state.
The system will be considered mechanically static
when the kinetic energy of the pile falls below $1.6^{-8}$J. This cut-off
value was deemed sufficient after examination of the system dynamics and its
kinetic energy evolution. 
The amplitude $A$ of the tap becomes the control parameter while $\omega$ remains
constant, equal to $\omega=134$hz. Although there is an open discussion~\cite{dijksman2009role,pugnaloni2008nonmonotonic,ludewig2008energetic,gago2015relevance}
about the proper parameter to characterize tap energy, following the
most standard usage, results will be reported as function of the dimensionless
acceleration $\Gamma=\omega^{2}\times A/g$, where $g$ is the
 acceleration due to gravity. 

Because of gravity, density is not uniform along the height of the
system~\cite{Nowak98,mehta2009heterogeneities,gago2015relevance}.
For this reason, the analysis is performed for three narrow horizontal slices at different heights of the silo, at  $z=3-4$cm, $z=4.5-5.5$cm, and $z=6-7$cm. To avoid wall effects, only the $2$cm diameter inner region is considered. Then, each slice contains of the order of $3500$ particles. 
Apart from the quantitative differences imposed by gravity, all regions behave consistently, and, for simplicity, results corresponding only to the central slice ($z=4.5-5.5$cm) are shown in this letter.
Analysis corresponding to the remaining slices can be found in the supplementary materials.

Local ``grain densities'' are obtained by dividing grain
volume by their corresponding Voronoi volume, obtained using voro++~\cite{Rycroft2009} open source 
software. The reported density (or packing fraction) $\phi$ for the system is then obtained by averaging local grain densities of the grains over the observed region. Ten independent realizations are
performed for each experiment.

Using a step-wise protocol inspired by the Chicago 
experiments, we obtain a ``reversible branch''~\cite{Nowak98}, that will serve as a reference for the annealing simulations.
It follows a step-wise ramp-down of the tap acceleration $\Gamma$, initiated at a
high value such as to avoid the ``non-reversible'' density branch~\cite{pugnaloni2008nonmonotonic}. 
Starting with a loosely poured configuration at $\Gamma=12$,   a set of $150$ taps are applied at each 
$\Gamma$, followed by a decreased with a wide step, $-\Delta\Gamma$, for another set
of $150$ taps, and so on. Only the density for the last $100$ taps
at each $\Gamma$ is averaged over, meant to avoid transients. 
Fig.\ \ref{fig:Gammadot} shows the packing fraction $\phi$ for the middle part of the system obtained in this manner as a function of $\Gamma$,
averaged over ten independent realizations.

Our annealing simulations, also shown in Fig.\ \ref{fig:Gammadot},
employ instead a protocol of \emph{gradual} changes after each tap,
reducing $\Gamma$ minutely at a fixed $-\dot{\Gamma}$. 
In this annealing protocol, we start from the stationary state created
by the previous protocol at a high value of $\Gamma_{{\rm high}}=12$.
Then, $\Gamma$ is reduced by $-\dot{\Gamma}$ each tap until reaching well below $\Gamma<1$, 
where the density plateaus into a terminal density
$\left\langle \phi_{\Gamma\to0}\right\rangle $ that depends on $-\dot{\Gamma}$. We repeat these simulations for various $-\dot{\Gamma}$.  In the inset of Fig.\ \ref{fig:Gammadot}
we plot $\left\langle \phi_{\Gamma\to0}\right\rangle $ as a function
of $-\dot{\Gamma}$. It varies rapidly, albeit systematically, with
a sub-linear exponent and an apparently much lower ``ideal'' equilibrium
density $\left\langle \phi{}_{\Gamma\to0}\right\rangle$ for $-\dot{\Gamma}\to0$, although both are difficult to determine from
fitting this data.

The densities during gradual annealing as well as during the stepped
Chicago protocol track each other closely for high accelerations $\Gamma$ and,
as a function of is ramp-down with $-\dot{\Gamma}$, they start deviating
from each other at lower $\Gamma$. This suggests a transition into
a glassy, non-equilibrium regime. This behavior closely resembles
glass transitions found in corresponding studies of polymers and many
other glasses~\cite{Hutchinson1995,Debenedetti01,Zhu13,roth2016polymer,Gedde19},
and it is consistent with previous studies finding ``non-ergocdicity''
in these systems at low values of $\Gamma$~\cite{paillusson2012probing,gago2017ergodic}. In particular, at very low $-\dot{\Gamma}\to0$,
 densities are reached that are above the ostensibly reversible branch,
 projecting an even more dense ideal protocol in the limit than is
 found in the Chicago pile~\cite{Nowak98,Nowak97}.  [Note that we plot increasing density downward to conform with most of the previous literature on annealing of glasses, which concerns the loss of free energy, corresponding here to a loss of free volume and gravitational potential energy that amounts to the complement, $1-\phi$, of the density.] The entire annealing
behavior for a granular pile, as exhibited in Fig.\ \ref{fig:Gammadot},
clearly warrants more detailed investigation in the future. Here,
we merely note the existence of an actual glass-like transition.

For the remainder of this Letter, we focus on the aging protocol indicated
by dashed arrows in Fig.\ \ref{fig:Gammadot}. The aging dynamics
is induced by a hard quench ($-\dot{\Gamma}=\infty$) from an equilibrated 
state at $\Gamma_{{\rm high}}=12$ into the glassy
regime at $\Gamma_{q}=1.3$, just above $\Gamma=1$, below which
the acceleration in a tap is insufficient to excite further changes.
At $\Gamma_{q}$, the acceleration of the pile is high enough that
relaxation can proceed, but sufficiently low inside the glassy regime
to avoid equilibration on any experimentally reasonable timeline.
We explore the aging dynamics for a sequence of $2^{12}$ taps.

\begin{figure}
\centering \includegraphics[width=1\columnwidth]{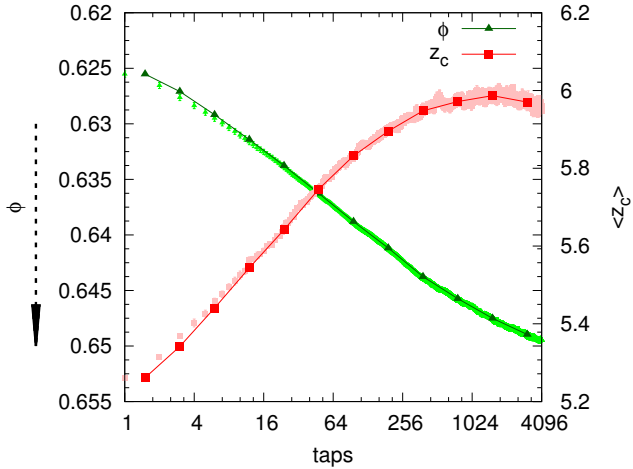} 
\caption{Average density $\phi$ of the granular
pile (left axis) and average number of contacts, $\left\langle z_{c}\right\rangle$
(right axis), as a function of number of taps, after a quench from $\Gamma_{{\rm high}}=12$ to $\Gamma_{q}=1.3$, averaged over 10 runs.
Solid symbols show the log-binned average for each variable.}
\label{fig:agingdensity} 
\end{figure}

In Fig.\ \ref{fig:agingdensity}, we plot the evolution of the average
density $\phi$ within the observation region as a function of taps.
Consistent with Ref.~\cite{Nowak98},
$\phi$ increases logarithmically with the number of taps during aging.  
This is also analogous to state variables in many  other glassy (thermal) systems~\cite{Hutchinson1995,Amir2012,roth2016polymer,Yunker09,Robe18}.
As minute amounts of free volume and gravitational
potential energy get dissipated via friction and collisions, we also
find a commensurate increase in the average number of supporting contacts
$\left\langle z_{c}\right\rangle$ holding each grain in position
after a tap (red squares in Fig.~\ref{fig:agingdensity}).
% In fact, examining individual runs of the aging protocol (exemplified
% in the inset of Fig.\ \ref{fig:agingdensity}) demonstrates that
% fluctuations in both measures largely coincide, suggesting a causal
% relation. In fact, Pearson's correlation coefficient $r$ of both
% time series, $\phi(t_{w})$ and $z_{c}(t_{w})$, when averaged of
% the ten runs, is rather uniform in every time interval, and we obtain
% a value of $r\approx0.57$ after averaging over all times. Thus, the
% events leading to increasing contacts between grains are essential
% to the irreversible decrease in energy.

\begin{figure}
\centering \includegraphics[width=1\columnwidth]{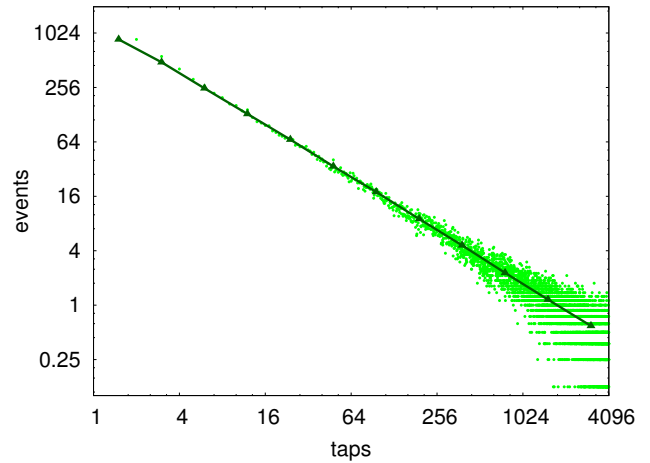}
\caption{Number of new record events per unit time  (tap), where an event is accounted when a particle increases its coordination number from $z_{c}\to z_{c}+1$ for the first time, averaged over ten runs.
The full symbols show the log-binned averaged. }
\label{fig:changingneighbors} 
\end{figure}

Identifying and counting events that facilitate such irreversible
changes is far from straightforward. We define as such an event the moment at which a given grain for the
first time increases its coordination number
from $z_{c}\to z_{c}+1$. Although its coordination number may still decrease
occasionally, achieving a new record in contacts for the first time
signifies an irreversible structural change in the contact network
of that grain and the forces holding it in place. We argue that such
an event permanently modifies the energetic landscape. Fig.\ \ref{fig:changingneighbors}
shows the decelerating rate at which these record-breaking events
occur, essentially hyperbolically, $\sim1/t$, over four decades
as a function of time $t$ since the quench. Assuming
that each ``event'' dissipates approximately an equal amount of
potential energy, then the accumulated decline in potential energy
-- and correspondingly, the increase in density -- as a logarithm
in time follows quite naturally as the integral of this rate, as we
will argue below.

\begin{figure}
\centering \includegraphics[width=1\columnwidth]{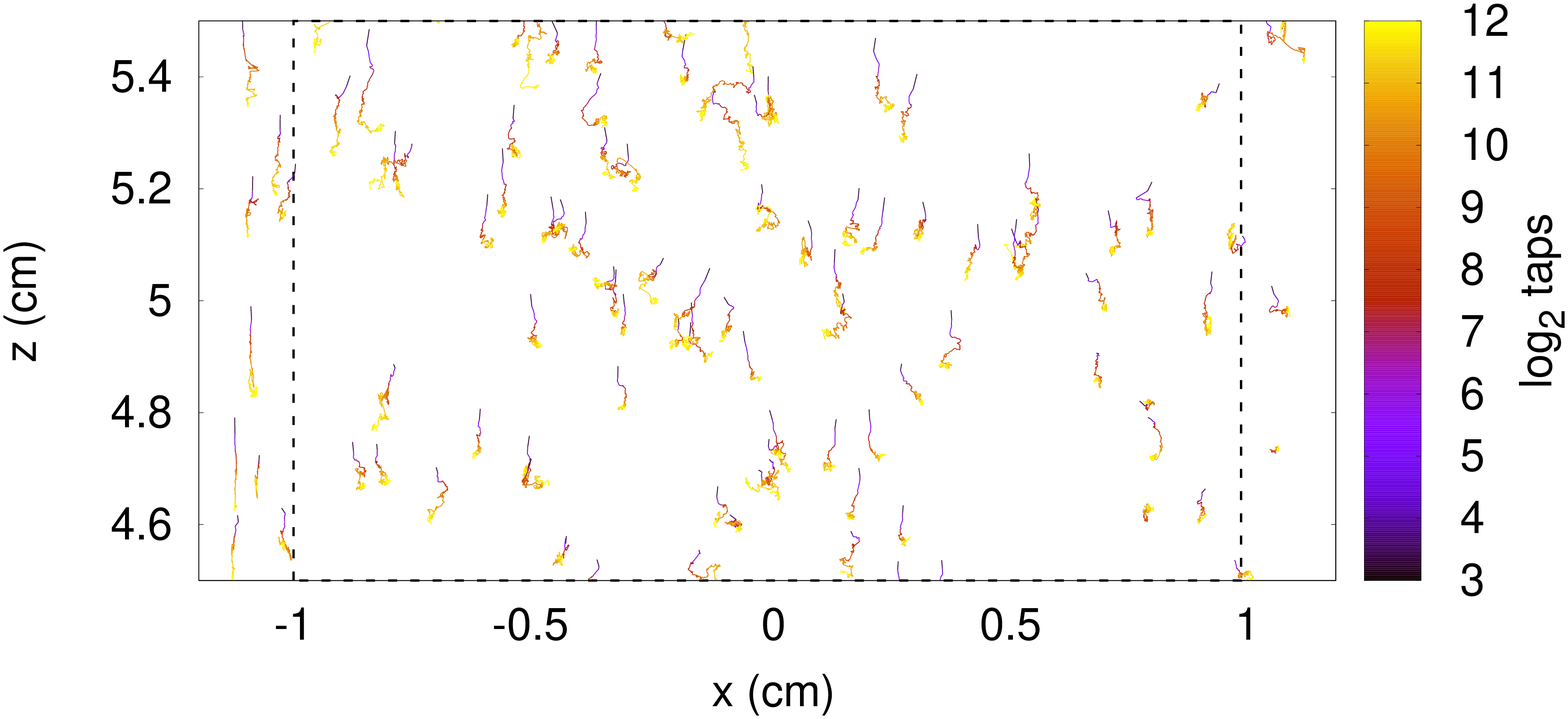}
\caption{Tracking of the displacement for a few randomly selected particles in the observed region (dashed box) inside the pile. The color-coding is logarithmic in time (taps) and indicates a decelerating downward drift along increasingly rugged (intermittent) trajectories. }
\label{fig:track} 
\end{figure}

It proves prohibitive to related such a topological change in the
contact network to an immediate, adjacent increase in density. Due
to gravity, discrete topological changes in one place may often lead
to a net increase in local density spread widely and much farther
upstream. However, the accumulative effect of these events reveals
itself microscopically in the net downward drift of all grains individually, as Fig.~\ref{fig:track} shows. Unlike for colloidal systems in the absence of gravity, 
where diffusion drives relaxation and dissipation
of free energy~\cite{Yunker09,Robe16}, no discernabile mean-square
displacement within the horizontal plane is present in the granular pile. Instead, particles predominantly drift downward in intermittent steps, ever more slowly, as those tracks in Fig.~\ref{fig:track} indicate. Such downward displacements signify the loss of free energy, the main
mechanism by which this system relaxes and density increases.

\begin{figure}
\centering \includegraphics[width=1\columnwidth]{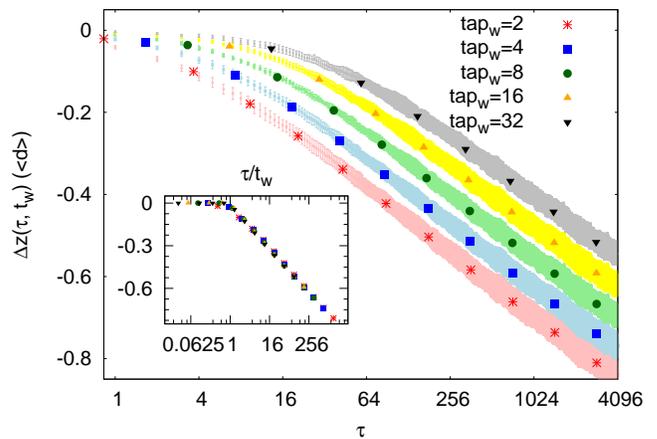}

\caption{Plot of the downward displacement averaged over all grains within
the observation region as a function of lag-time $\tau=t-t_{w},$
relative to each grain's position at time $t_{w}$ after the quench.
\emph{Inset:} Collapse of the data in the main panel when plotted
as function of $\tau/t_{w}$.}

\label{fig:displacement} 
\end{figure}

To examine the history of particle trajectories in more detail, we study the downward drift as a function of \emph{two} times, namely the displacement achieved at $t$ taps relative to the height attained after a number of $t_w$ taps beyond the quench. In equilibrium, such a measure would be translationally invariant and merely depend on the difference $t-t_w$. An aging system is characterized by its history dependence via the "waiting time" $t_w$. The overall decelerating dynamics is reflected in the fact that, the later $t_w$, the less particles manage to advance during the subsequent "lag-time" $\tau=t-t_w$. However, as Fig.~\ref{fig:displacement} shows, such waiting-time dependence assumes a very specific form for the granular pile, similar to other glassy systems~\cite{Dupuis2005,Rodriguez03,Robe16}. There, we plot the average downward displacement
for all grains in the observed region as a function of lag-time $\tau$ for a geometric progression of waiting times. Notably,
the data collapses reasonably well when time is simply rescaled by the age
of the quench process itself and plotted as function of $\tau/t_{w}=\frac{t}{t_{w}}-1$,
as the inset shows \footnote{In some experiments, such as for thermo-remanent relaxation in spin glasses~\cite{Dupuis2005,Rodriguez03}, it is necessary to actively perturb the system at a time $t_w$ to elicit as measurable response at a later time $t$. However, in simulations where one controls the state of every variable at $t_w$ and $t$, such perturbations are unnecessary.}.

Remarkably, we thus find that a tapped granular pile exhibits an
aging phenomenology, after being driven out of equilibrium by a hard
quench, akin to many other thermal systems, not only macroscopically~\cite{Nowak98}
but also in the statistics of local events facilitating relaxation.
The statistics of these events closely resemble a log-Poisson process,
which is at the heart of a universal description -- essential for
explaining the ubiquity of this phenomenology across so many materials
-- in terms of record dynamics~\cite{Sibani03,Anderson04,BoSi09,SJ13,Boettcher18b}.
These events are enabled, in fact, by statistically independent, record-sized
fluctuations that evolve the dynamics irreversibly and in ever more
rare increments. Hence, the instantaneous rate $\lambda(t)$ of irreversible
events as a function of time ($t$, in taps) should follow a simple statistic of records, for which $\lambda\sim1/t$,
the chance that the next number drawn randomly out of $t$ is
the largest. Activity decelerates as an aging system ``stiffens''
and a new record event is needed to evolve further. Unlike for a Poisson
process with a constant rate, the average number of intermittent events
in an interval $\left(t_{w},t\right]$ for a log-Poisson process is
\begin{equation}
\left\langle n_{I}\left(t,t_{w}\right)\right\rangle =\int_{t_{w}}^{t}\lambda\left(t^{\prime}\right)\,dt^{\prime}\sim\ln\left(t/t_{w}\right),\label{eq:LogPoisson}
\end{equation}
entailing an explicit memory of the age $t_{w}$.

Indeed, we find that the rate of meaningful events in stiffening a
grain's contact network follows such a statistics of records, as shown
in Fig.\ \ref{fig:changingneighbors}. Then, according to Eq.\ (\ref{eq:LogPoisson}),
any two-time functions, such as the displacements in Fig.\ \ref{fig:displacement},
become \emph{subordinate}~\cite{SJ13,Robe18} to this process:
$C\left(t,t_{w}\right)=C\left[n_{I}\right]=C\left(t/t_{w}\right)$,
implying both, activated displacements that increase logarithmic in time as well as the data collapse in the inset. In particular, the logarithmic
behavior in time of one-time macroscopic observables, such as the
density of the system in Fig.\ \ref{fig:agingdensity}, follows equally
from Eq.\ (\ref{eq:LogPoisson}). Previously, those predictions already
have been verified in experiments and simulations with colloids~\cite{Yunker09,Robe16,Robe18}
and spin glasses~\cite{Sibani06a,Sibani18}.

In conclusion, we have found that a granular pile has similar relaxation
behavior after a quench compared with other glassy materials. While
there has been much discussion concerning the possibility of describing
athermal granular systems in terms of an equilibrium thermodynamics~\cite{edwards1989theory,blumenfeld2009granular,pugnaloni2010towards,henkes2007entropy},
our study shows that, with regard to the aging phenomenology, granular
piles belong in a wide class of glassy materials that exhibit universal
behavior out of equilibrium. Our data is consistent with predictions
based on a log-Poisson process, in which increasingly rare, record-sized
fluctuations provide the activation to move a glassy system from one
meta-stable state to the next, marginally more stable state and irreversibly
exspell free energy. As the origin of such fluctuations, either from
thermal or from athermal driving, is irrelevant for the existence
of records, such a theory is well-suited to account for the ubiquity
of this phenomenology across many materials.

In the process, we also found that during the annealing of a granular
pile, at finite but very slow, fixed damping, the observed relaxation
trajectories undergo a glass transition similar to polymer glasses, as shown in Fig.~\ref{fig:Gammadot}.
Beyond that transition, the system falls out of equilibrium and settles
into an ensemble of meta-stable states whose average energy (i.e.,
its density), varies systematically with the annealing speed. The
variable dependence of macroscopic observables on annealing speed
has not been noted previously and contradicts established assumptions
about granular piles and, thus, warrants future investigation.

We are happy to acknowledge fruitful discussions about this project
with P. Sibani. These simulation were performed at the Imperial College Research Computing Service (see DOI: 10.14469/hpc/2232). 

\bibliographystyle{apsrev4-1}
\bibliography{/Users/sboettc/Boettcher}

\begin{thebibliography}{46}%
\makeatletter
\providecommand \@ifxundefined [1]{%
 \@ifx{#1\undefined}
}%
\providecommand \@ifnum [1]{%
 \ifnum #1\expandafter \@firstoftwo
 \else \expandafter \@secondoftwo
 \fi
}%
\providecommand \@ifx [1]{%
 \ifx #1\expandafter \@firstoftwo
 \else \expandafter \@secondoftwo
 \fi
}%
\providecommand \natexlab [1]{#1}%
\providecommand \enquote  [1]{``#1''}%
\providecommand \bibnamefont  [1]{#1}%
\providecommand \bibfnamefont [1]{#1}%
\providecommand \citenamefont [1]{#1}%
\providecommand \href@noop [0]{\@secondoftwo}%
\providecommand \href [0]{\begingroup \@sanitize@url \@href}%
\providecommand \@href[1]{\@@startlink{#1}\@@href}%
\providecommand \@@href[1]{\endgroup#1\@@endlink}%
\providecommand \@sanitize@url [0]{\catcode `\\12\catcode `\$12\catcode
  `\&12\catcode `\#12\catcode `\^12\catcode `\_12\catcode `\%12\relax}%
\providecommand \@@startlink[1]{}%
\providecommand \@@endlink[0]{}%
\providecommand \url  [0]{\begingroup\@sanitize@url \@url }%
\providecommand \@url [1]{\endgroup\@href {#1}{\urlprefix }}%
\providecommand \urlprefix  [0]{URL }%
\providecommand \Eprint [0]{\href }%
\providecommand \doibase [0]{http://dx.doi.org/}%
\providecommand \selectlanguage [0]{\@gobble}%
\providecommand \bibinfo  [0]{\@secondoftwo}%
\providecommand \bibfield  [0]{\@secondoftwo}%
\providecommand \translation [1]{[#1]}%
\providecommand \BibitemOpen [0]{}%
\providecommand \bibitemStop [0]{}%
\providecommand \bibitemNoStop [0]{.\EOS\space}%
\providecommand \EOS [0]{\spacefactor3000\relax}%
\providecommand \BibitemShut  [1]{\csname bibitem#1\endcsname}%
\let\auto@bib@innerbib\@empty
%</preamble>
\bibitem [{\citenamefont {{Liu}}\ and\ \citenamefont {{Nagel}}(1998)}]{Liu98}%
  \BibitemOpen
  \bibfield  {author} {\bibinfo {author} {\bibfnamefont {A.~J.}\ \bibnamefont
  {{Liu}}}\ and\ \bibinfo {author} {\bibfnamefont {S.~R.}\ \bibnamefont
  {{Nagel}}},\ }\href {\doibase 10.1038/23819} {\bibfield  {journal} {\bibinfo
  {journal} {Nature}\ }\textbf {\bibinfo {volume} {396}},\ \bibinfo {pages}
  {21} (\bibinfo {year} {1998})}\BibitemShut {NoStop}%
\bibitem [{\citenamefont {Buisson}\ \emph {et~al.}(2003)\citenamefont
  {Buisson}, \citenamefont {Bellon},\ and\ \citenamefont
  {Ciliberto}}]{Buisson03}%
  \BibitemOpen
  \bibfield  {author} {\bibinfo {author} {\bibfnamefont {L.}~\bibnamefont
  {Buisson}}, \bibinfo {author} {\bibfnamefont {L.}~\bibnamefont {Bellon}}, \
  and\ \bibinfo {author} {\bibfnamefont {S.}~\bibnamefont {Ciliberto}},\
  }\href@noop {} {\bibfield  {journal} {\bibinfo  {journal} {J. Phys. Cond.
  Mat.}\ }\textbf {\bibinfo {volume} {15}},\ \bibinfo {pages} {S1163} (\bibinfo
  {year} {2003})}\BibitemShut {NoStop}%
\bibitem [{\citenamefont {{L.~Buisson, S.~Ciliberto and A.
  Garciamartin}}(2003)}]{Buisson03a}%
  \BibitemOpen
  \bibfield  {author} {\bibinfo {author} {\bibnamefont {{L.~Buisson,
  S.~Ciliberto and A. Garciamartin}}},\ }\href@noop {} {\bibfield  {journal}
  {\bibinfo  {journal} {Europhys. Lett.}\ }\textbf {\bibinfo {volume} {63}},\
  \bibinfo {pages} {603} (\bibinfo {year} {2003})}\BibitemShut {NoStop}%
\bibitem [{\citenamefont {Bissig}\ \emph {et~al.}(2003)\citenamefont {Bissig},
  \citenamefont {Romer}, \citenamefont {Cipelletti}, \citenamefont {Trappe},\
  and\ \citenamefont {Schurtenberger}}]{Bissig03}%
  \BibitemOpen
  \bibfield  {author} {\bibinfo {author} {\bibfnamefont {H.}~\bibnamefont
  {Bissig}}, \bibinfo {author} {\bibfnamefont {S.}~\bibnamefont {Romer}},
  \bibinfo {author} {\bibfnamefont {L.}~\bibnamefont {Cipelletti}}, \bibinfo
  {author} {\bibfnamefont {V.}~\bibnamefont {Trappe}}, \ and\ \bibinfo {author}
  {\bibfnamefont {P.}~\bibnamefont {Schurtenberger}},\ }\href {\doibase
  10.1039/b211806h} {\bibfield  {journal} {\bibinfo  {journal} {Phys. Chem.
  Comm.}\ }\textbf {\bibinfo {volume} {6}},\ \bibinfo {pages} {21} (\bibinfo
  {year} {2003})}\BibitemShut {NoStop}%
\bibitem [{\citenamefont {Yunker}\ \emph {et~al.}(2009)\citenamefont {Yunker},
  \citenamefont {Zhang}, \citenamefont {Aptowicz},\ and\ \citenamefont
  {Yodh}}]{Yunker09}%
  \BibitemOpen
  \bibfield  {author} {\bibinfo {author} {\bibfnamefont {P.}~\bibnamefont
  {Yunker}}, \bibinfo {author} {\bibfnamefont {Z.}~\bibnamefont {Zhang}},
  \bibinfo {author} {\bibfnamefont {K.~B.}\ \bibnamefont {Aptowicz}}, \ and\
  \bibinfo {author} {\bibfnamefont {A.~G.}\ \bibnamefont {Yodh}},\ }\href@noop
  {} {\bibfield  {journal} {\bibinfo  {journal} {Phys. Rev. Lett.}\ }\textbf
  {\bibinfo {volume} {103}},\ \bibinfo {pages} {115701} (\bibinfo {year}
  {2009})}\BibitemShut {NoStop}%
\bibitem [{\citenamefont {Kajiya}\ \emph {et~al.}(2013)\citenamefont {Kajiya},
  \citenamefont {Narita}, \citenamefont {Schmitt}, \citenamefont {Lequeuxa},\
  and\ \citenamefont {Talini}}]{Kajiya13}%
  \BibitemOpen
  \bibfield  {author} {\bibinfo {author} {\bibfnamefont {T.}~\bibnamefont
  {Kajiya}}, \bibinfo {author} {\bibfnamefont {T.}~\bibnamefont {Narita}},
  \bibinfo {author} {\bibfnamefont {V.}~\bibnamefont {Schmitt}}, \bibinfo
  {author} {\bibfnamefont {F.}~\bibnamefont {Lequeuxa}}, \ and\ \bibinfo
  {author} {\bibfnamefont {L.}~\bibnamefont {Talini}},\ }\href@noop {}
  {\bibfield  {journal} {\bibinfo  {journal} {Soft Matter}\ }\textbf {\bibinfo
  {volume} {9}},\ \bibinfo {pages} {11129} (\bibinfo {year}
  {2013})}\BibitemShut {NoStop}%
\bibitem [{\citenamefont {Zargar}\ \emph {et~al.}(2013)\citenamefont {Zargar},
  \citenamefont {Nienhuis}, \citenamefont {Schall},\ and\ \citenamefont
  {Bonn}}]{Zargar13}%
  \BibitemOpen
  \bibfield  {author} {\bibinfo {author} {\bibfnamefont {R.}~\bibnamefont
  {Zargar}}, \bibinfo {author} {\bibfnamefont {B.}~\bibnamefont {Nienhuis}},
  \bibinfo {author} {\bibfnamefont {P.}~\bibnamefont {Schall}}, \ and\ \bibinfo
  {author} {\bibfnamefont {D.}~\bibnamefont {Bonn}},\ }\href@noop {} {\bibfield
   {journal} {\bibinfo  {journal} {Phys. Rev. Lett.}\ }\textbf {\bibinfo
  {volume} {110}},\ \bibinfo {pages} {258301} (\bibinfo {year}
  {2013})}\BibitemShut {NoStop}%
\bibitem [{\citenamefont {Yanagishima}\ \emph {et~al.}(2017)\citenamefont
  {Yanagishima}, \citenamefont {Russo},\ and\ \citenamefont
  {Tanaka}}]{Tanaka17}%
  \BibitemOpen
  \bibfield  {author} {\bibinfo {author} {\bibfnamefont {T.}~\bibnamefont
  {Yanagishima}}, \bibinfo {author} {\bibfnamefont {J.}~\bibnamefont {Russo}},
  \ and\ \bibinfo {author} {\bibfnamefont {H.}~\bibnamefont {Tanaka}},\
  }\href@noop {} {\bibfield  {journal} {\bibinfo  {journal} {Nature
  Communications}\ }\textbf {\bibinfo {volume} {8}},\ \bibinfo {pages} {15954}
  (\bibinfo {year} {2017})}\BibitemShut {NoStop}%
\bibitem [{\citenamefont {Struik}(1978)}]{Struik78}%
  \BibitemOpen
  \bibfield  {author} {\bibinfo {author} {\bibfnamefont {L.}~\bibnamefont
  {Struik}},\ }\href@noop {} {\emph {\bibinfo {title} {Physical aging in
  amorphous polymers and other materials}}}\ (\bibinfo  {publisher} {Elsevier
  Science Ltd},\ \bibinfo {address} {New York},\ \bibinfo {year}
  {1978})\BibitemShut {NoStop}%
\bibitem [{\citenamefont {{P. Sibani, G.F. Rodriguez and G.G.
  Kenning}}(2006)}]{Sibani06a}%
  \BibitemOpen
  \bibfield  {author} {\bibinfo {author} {\bibnamefont {{P. Sibani, G.F.
  Rodriguez and G.G. Kenning}}},\ }\href@noop {} {\bibfield  {journal}
  {\bibinfo  {journal} {Phys. Rev. B}\ }\textbf {\bibinfo {volume} {74}},\
  \bibinfo {pages} {224407} (\bibinfo {year} {2006})}\BibitemShut {NoStop}%
\bibitem [{\citenamefont {Oliveira}\ \emph {et~al.}(2005)\citenamefont
  {Oliveira}, \citenamefont {Jensen}, \citenamefont {Nicodemi},\ and\
  \citenamefont {Sibani}}]{Oliveira05}%
  \BibitemOpen
  \bibfield  {author} {\bibinfo {author} {\bibfnamefont {L.~P.}\ \bibnamefont
  {Oliveira}}, \bibinfo {author} {\bibfnamefont {H.~J.}\ \bibnamefont
  {Jensen}}, \bibinfo {author} {\bibfnamefont {M.}~\bibnamefont {Nicodemi}}, \
  and\ \bibinfo {author} {\bibfnamefont {P.}~\bibnamefont {Sibani}},\
  }\href@noop {} {\bibfield  {journal} {\bibinfo  {journal} {Phys. Rev. B}\
  }\textbf {\bibinfo {volume} {71}},\ \bibinfo {pages} {104526} (\bibinfo
  {year} {2005})}\BibitemShut {NoStop}%
\bibitem [{\citenamefont {Boettcher}\ and\ \citenamefont
  {Sibani}(2011)}]{BoSi09}%
  \BibitemOpen
  \bibfield  {author} {\bibinfo {author} {\bibfnamefont {S.}~\bibnamefont
  {Boettcher}}\ and\ \bibinfo {author} {\bibfnamefont {P.}~\bibnamefont
  {Sibani}},\ }\href@noop {} {\bibfield  {journal} {\bibinfo  {journal} {J.
  Phys.: Condens. Matter}\ }\textbf {\bibinfo {volume} {23}},\ \bibinfo {pages}
  {065103} (\bibinfo {year} {2011})}\BibitemShut {NoStop}%
\bibitem [{\citenamefont {Robe}\ \emph {et~al.}(2016)\citenamefont {Robe},
  \citenamefont {Boettcher}, \citenamefont {Sibani},\ and\ \citenamefont
  {Yunker}}]{Robe16}%
  \BibitemOpen
  \bibfield  {author} {\bibinfo {author} {\bibfnamefont {D.~M.}\ \bibnamefont
  {Robe}}, \bibinfo {author} {\bibfnamefont {S.}~\bibnamefont {Boettcher}},
  \bibinfo {author} {\bibfnamefont {P.}~\bibnamefont {Sibani}}, \ and\ \bibinfo
  {author} {\bibfnamefont {P.}~\bibnamefont {Yunker}},\ }\href@noop {}
  {\bibfield  {journal} {\bibinfo  {journal} {EuroPhys. Lett.}\ }\textbf
  {\bibinfo {volume} {116}},\ \bibinfo {pages} {38003} (\bibinfo {year}
  {2016})}\BibitemShut {NoStop}%
\bibitem [{\citenamefont {Nowak}\ \emph {et~al.}(1998)\citenamefont {Nowak},
  \citenamefont {Knight}, \citenamefont {Ben-Naim}, \citenamefont {Jaeger},\
  and\ \citenamefont {Nagel}}]{Nowak98}%
  \BibitemOpen
  \bibfield  {author} {\bibinfo {author} {\bibfnamefont {E.~R.}\ \bibnamefont
  {Nowak}}, \bibinfo {author} {\bibfnamefont {J.~B.}\ \bibnamefont {Knight}},
  \bibinfo {author} {\bibfnamefont {E.}~\bibnamefont {Ben-Naim}}, \bibinfo
  {author} {\bibfnamefont {H.~M.}\ \bibnamefont {Jaeger}}, \ and\ \bibinfo
  {author} {\bibfnamefont {S.~R.}\ \bibnamefont {Nagel}},\ }\href@noop {}
  {\bibfield  {journal} {\bibinfo  {journal} {Phys. Rev. E}\ }\textbf {\bibinfo
  {volume} {57}},\ \bibinfo {pages} {1971} (\bibinfo {year}
  {1998})}\BibitemShut {NoStop}%
\bibitem [{\citenamefont {Nowak}\ \emph {et~al.}(1997)\citenamefont {Nowak},
  \citenamefont {Knight}, \citenamefont {Povinelli}, \citenamefont {Jaeger},\
  and\ \citenamefont {Nagel}}]{Nowak97}%
  \BibitemOpen
  \bibfield  {author} {\bibinfo {author} {\bibfnamefont {E.}~\bibnamefont
  {Nowak}}, \bibinfo {author} {\bibfnamefont {J.}~\bibnamefont {Knight}},
  \bibinfo {author} {\bibfnamefont {M.}~\bibnamefont {Povinelli}}, \bibinfo
  {author} {\bibfnamefont {H.}~\bibnamefont {Jaeger}}, \ and\ \bibinfo {author}
  {\bibfnamefont {S.}~\bibnamefont {Nagel}},\ }\href {\doibase
  10.1016/s0032-5910(97)03291-9} {\bibfield  {journal} {\bibinfo  {journal}
  {Powder Technology}\ }\textbf {\bibinfo {volume} {94}},\ \bibinfo {pages}
  {79} (\bibinfo {year} {1997})}\BibitemShut {NoStop}%
\bibitem [{\citenamefont {Hutchinson}(1995)}]{Hutchinson1995}%
  \BibitemOpen
  \bibfield  {author} {\bibinfo {author} {\bibfnamefont {J.~M.}\ \bibnamefont
  {Hutchinson}},\ }\href {\doibase 10.1016/0079-6700(94)00001-i} {\bibfield
  {journal} {\bibinfo  {journal} {Progress in Polymer Science}\ }\textbf
  {\bibinfo {volume} {20}},\ \bibinfo {pages} {703} (\bibinfo {year}
  {1995})}\BibitemShut {NoStop}%
\bibitem [{\citenamefont {Amir}\ \emph {et~al.}(2012)\citenamefont {Amir},
  \citenamefont {Oreg},\ and\ \citenamefont {Imry}}]{Amir2012}%
  \BibitemOpen
  \bibfield  {author} {\bibinfo {author} {\bibfnamefont {A.}~\bibnamefont
  {Amir}}, \bibinfo {author} {\bibfnamefont {Y.}~\bibnamefont {Oreg}}, \ and\
  \bibinfo {author} {\bibfnamefont {Y.}~\bibnamefont {Imry}},\ }\href {\doibase
  10.1073/pnas.1120147109} {\bibfield  {journal} {\bibinfo  {journal}
  {Proceedings of the National Academy of Sciences}\ }\textbf {\bibinfo
  {volume} {109}},\ \bibinfo {pages} {1850} (\bibinfo {year}
  {2012})}\BibitemShut {NoStop}%
\bibitem [{\citenamefont {Roth}(2016)}]{roth2016polymer}%
  \BibitemOpen
  \bibfield  {author} {\bibinfo {author} {\bibfnamefont {C.}~\bibnamefont
  {Roth}},\ }\href {https://books.google.com/books?id=SQsNDgAAQBAJ} {\emph
  {\bibinfo {title} {Polymer Glasses}}}\ (\bibinfo  {publisher} {CRC Press},\
  \bibinfo {year} {2016})\BibitemShut {NoStop}%
\bibitem [{\citenamefont {Ben-Naim}\ \emph {et~al.}(1998)\citenamefont
  {Ben-Naim}, \citenamefont {Knight}, \citenamefont {Nowak}, \citenamefont
  {Jaeger},\ and\ \citenamefont {Nagel}}]{BenNaim98}%
  \BibitemOpen
  \bibfield  {author} {\bibinfo {author} {\bibfnamefont {E.}~\bibnamefont
  {Ben-Naim}}, \bibinfo {author} {\bibfnamefont {J.}~\bibnamefont {Knight}},
  \bibinfo {author} {\bibfnamefont {E.}~\bibnamefont {Nowak}}, \bibinfo
  {author} {\bibfnamefont {H.}~\bibnamefont {Jaeger}}, \ and\ \bibinfo {author}
  {\bibfnamefont {S.}~\bibnamefont {Nagel}},\ }\href@noop {} {\bibfield
  {journal} {\bibinfo  {journal} {Physica D: Nonlinear Phenomena}\ }\textbf
  {\bibinfo {volume} {123}},\ \bibinfo {pages} {380 } (\bibinfo {year}
  {1998})}\BibitemShut {NoStop}%
\bibitem [{\citenamefont {Sibani}\ and\ \citenamefont
  {Boettcher}(2016)}]{Sibani16}%
  \BibitemOpen
  \bibfield  {author} {\bibinfo {author} {\bibfnamefont {P.}~\bibnamefont
  {Sibani}}\ and\ \bibinfo {author} {\bibfnamefont {S.}~\bibnamefont
  {Boettcher}},\ }\href {\doibase 10.1103/PhysRevE.93.062141} {\bibfield
  {journal} {\bibinfo  {journal} {Phys. Rev. E}\ }\textbf {\bibinfo {volume}
  {93}},\ \bibinfo {pages} {062141} (\bibinfo {year} {2016})}\BibitemShut
  {NoStop}%
\bibitem [{\citenamefont {Gago}\ \emph {et~al.}(2016)\citenamefont {Gago},
  \citenamefont {Maza},\ and\ \citenamefont {Pugnaloni}}]{gago2017ergodic}%
  \BibitemOpen
  \bibfield  {author} {\bibinfo {author} {\bibfnamefont {P.~A.}\ \bibnamefont
  {Gago}}, \bibinfo {author} {\bibfnamefont {D.}~\bibnamefont {Maza}}, \ and\
  \bibinfo {author} {\bibfnamefont {L.~A.}\ \bibnamefont {Pugnaloni}},\ }\href
  {\doibase 10.4279/pip.080001} {\bibfield  {journal} {\bibinfo  {journal}
  {Papers in Physics}\ }\textbf {\bibinfo {volume} {8}} (\bibinfo {year}
  {2016}),\ 10.4279/pip.080001}\BibitemShut {NoStop}%
\bibitem [{\citenamefont {Debenedetti}\ and\ \citenamefont
  {Stillinger}(2001)}]{Debenedetti01}%
  \BibitemOpen
  \bibfield  {author} {\bibinfo {author} {\bibfnamefont {P.~G.}\ \bibnamefont
  {Debenedetti}}\ and\ \bibinfo {author} {\bibfnamefont {F.~H.}\ \bibnamefont
  {Stillinger}},\ }\href {\doibase 10.1038/35065704} {\bibfield  {journal}
  {\bibinfo  {journal} {Nature}\ }\textbf {\bibinfo {volume} {410}},\ \bibinfo
  {pages} {259} (\bibinfo {year} {2001})}\BibitemShut {NoStop}%
\bibitem [{\citenamefont {Zhu}\ \emph {et~al.}(2013)\citenamefont {Zhu},
  \citenamefont {Wen}, \citenamefont {Wang}, \citenamefont {Xue}, \citenamefont
  {Zhao},\ and\ \citenamefont {Wang}}]{Zhu13}%
  \BibitemOpen
  \bibfield  {author} {\bibinfo {author} {\bibfnamefont {Z.~G.}\ \bibnamefont
  {Zhu}}, \bibinfo {author} {\bibfnamefont {P.}~\bibnamefont {Wen}}, \bibinfo
  {author} {\bibfnamefont {D.~P.}\ \bibnamefont {Wang}}, \bibinfo {author}
  {\bibfnamefont {R.~J.}\ \bibnamefont {Xue}}, \bibinfo {author} {\bibfnamefont
  {D.~Q.}\ \bibnamefont {Zhao}}, \ and\ \bibinfo {author} {\bibfnamefont
  {W.~H.}\ \bibnamefont {Wang}},\ }\href {\doibase 10.1063/1.4819484}
  {\bibfield  {journal} {\bibinfo  {journal} {Journal of Applied Physics}\
  }\textbf {\bibinfo {volume} {114}},\ \bibinfo {pages} {083512} (\bibinfo
  {year} {2013})}\BibitemShut {NoStop}%
\bibitem [{\citenamefont {Gedde}\ and\ \citenamefont
  {Hedenqvist}(2019)}]{Gedde19}%
  \BibitemOpen
  \bibfield  {author} {\bibinfo {author} {\bibfnamefont {U.~W.}\ \bibnamefont
  {Gedde}}\ and\ \bibinfo {author} {\bibfnamefont {M.~S.}\ \bibnamefont
  {Hedenqvist}},\ }\href {\doibase 10.1007/978-3-030-29794-7} {\emph {\bibinfo
  {title} {Fundamental Polymer Science}}}\ (\bibinfo  {publisher} {Springer
  International Publishing},\ \bibinfo {year} {2019})\BibitemShut {NoStop}%
\bibitem [{\citenamefont {Brilliantov}\ \emph {et~al.}(1996)\citenamefont
  {Brilliantov}, \citenamefont {Spahn}, \citenamefont {Hertzsch},\ and\
  \citenamefont {P{\"o}schel}}]{brilliantov1996model}%
  \BibitemOpen
  \bibfield  {author} {\bibinfo {author} {\bibfnamefont {N.~V.}\ \bibnamefont
  {Brilliantov}}, \bibinfo {author} {\bibfnamefont {F.}~\bibnamefont {Spahn}},
  \bibinfo {author} {\bibfnamefont {J.-M.}\ \bibnamefont {Hertzsch}}, \ and\
  \bibinfo {author} {\bibfnamefont {T.}~\bibnamefont {P{\"o}schel}},\
  }\href@noop {} {\bibfield  {journal} {\bibinfo  {journal} {Physical review
  E}\ }\textbf {\bibinfo {volume} {53}},\ \bibinfo {pages} {5382} (\bibinfo
  {year} {1996})}\BibitemShut {NoStop}%
\bibitem [{\citenamefont {Goniva}\ \emph {et~al.}(2012)\citenamefont {Goniva},
  \citenamefont {Kloss}, \citenamefont {Deen}, \citenamefont {Kuipers},\ and\
  \citenamefont {Pirker}}]{goniva2012influence}%
  \BibitemOpen
  \bibfield  {author} {\bibinfo {author} {\bibfnamefont {C.}~\bibnamefont
  {Goniva}}, \bibinfo {author} {\bibfnamefont {C.}~\bibnamefont {Kloss}},
  \bibinfo {author} {\bibfnamefont {N.~G.}\ \bibnamefont {Deen}}, \bibinfo
  {author} {\bibfnamefont {J.~A.}\ \bibnamefont {Kuipers}}, \ and\ \bibinfo
  {author} {\bibfnamefont {S.}~\bibnamefont {Pirker}},\ }\href@noop {}
  {\bibfield  {journal} {\bibinfo  {journal} {Particuology}\ }\textbf {\bibinfo
  {volume} {10}},\ \bibinfo {pages} {582} (\bibinfo {year} {2012})}\BibitemShut
  {NoStop}%
\bibitem [{\citenamefont {Dijksman}\ and\ \citenamefont {van
  Hecke}(2009)}]{dijksman2009role}%
  \BibitemOpen
  \bibfield  {author} {\bibinfo {author} {\bibfnamefont {J.~A.}\ \bibnamefont
  {Dijksman}}\ and\ \bibinfo {author} {\bibfnamefont {M.}~\bibnamefont {van
  Hecke}},\ }\href@noop {} {\bibfield  {journal} {\bibinfo  {journal} {EPL
  (Europhysics Letters)}\ }\textbf {\bibinfo {volume} {88}},\ \bibinfo {pages}
  {44001} (\bibinfo {year} {2009})}\BibitemShut {NoStop}%
\bibitem [{\citenamefont {Pugnaloni}\ \emph {et~al.}(2008)\citenamefont
  {Pugnaloni}, \citenamefont {Mizrahi}, \citenamefont {Carlevaro},\ and\
  \citenamefont {Vericat}}]{pugnaloni2008nonmonotonic}%
  \BibitemOpen
  \bibfield  {author} {\bibinfo {author} {\bibfnamefont {L.~A.}\ \bibnamefont
  {Pugnaloni}}, \bibinfo {author} {\bibfnamefont {M.}~\bibnamefont {Mizrahi}},
  \bibinfo {author} {\bibfnamefont {C.~M.}\ \bibnamefont {Carlevaro}}, \ and\
  \bibinfo {author} {\bibfnamefont {F.}~\bibnamefont {Vericat}},\ }\href@noop
  {} {\bibfield  {journal} {\bibinfo  {journal} {Physical Review E}\ }\textbf
  {\bibinfo {volume} {78}},\ \bibinfo {pages} {051305} (\bibinfo {year}
  {2008})}\BibitemShut {NoStop}%
\bibitem [{\citenamefont {Ludewig}\ \emph {et~al.}(2008)\citenamefont
  {Ludewig}, \citenamefont {Dorbolo}, \citenamefont {Gilet},\ and\
  \citenamefont {Vandewalle}}]{ludewig2008energetic}%
  \BibitemOpen
  \bibfield  {author} {\bibinfo {author} {\bibfnamefont {F.}~\bibnamefont
  {Ludewig}}, \bibinfo {author} {\bibfnamefont {S.}~\bibnamefont {Dorbolo}},
  \bibinfo {author} {\bibfnamefont {T.}~\bibnamefont {Gilet}}, \ and\ \bibinfo
  {author} {\bibfnamefont {N.}~\bibnamefont {Vandewalle}},\ }\href@noop {}
  {\bibfield  {journal} {\bibinfo  {journal} {EPL (Europhysics Letters)}\
  }\textbf {\bibinfo {volume} {84}},\ \bibinfo {pages} {44001} (\bibinfo {year}
  {2008})}\BibitemShut {NoStop}%
\bibitem [{\citenamefont {Gago}\ \emph {et~al.}(2015)\citenamefont {Gago},
  \citenamefont {Maza},\ and\ \citenamefont {Pugnaloni}}]{gago2015relevance}%
  \BibitemOpen
  \bibfield  {author} {\bibinfo {author} {\bibfnamefont {P.~A.}\ \bibnamefont
  {Gago}}, \bibinfo {author} {\bibfnamefont {D.}~\bibnamefont {Maza}}, \ and\
  \bibinfo {author} {\bibfnamefont {L.~A.}\ \bibnamefont {Pugnaloni}},\
  }\href@noop {} {\bibfield  {journal} {\bibinfo  {journal} {Physical Review
  E}\ }\textbf {\bibinfo {volume} {91}},\ \bibinfo {pages} {032207} (\bibinfo
  {year} {2015})}\BibitemShut {NoStop}%
\bibitem [{\citenamefont {Mehta}\ \emph {et~al.}(2009)\citenamefont {Mehta},
  \citenamefont {Barker},\ and\ \citenamefont
  {Luck}}]{mehta2009heterogeneities}%
  \BibitemOpen
  \bibfield  {author} {\bibinfo {author} {\bibfnamefont {A.}~\bibnamefont
  {Mehta}}, \bibinfo {author} {\bibfnamefont {G.~C.}\ \bibnamefont {Barker}}, \
  and\ \bibinfo {author} {\bibfnamefont {J.-M.}\ \bibnamefont {Luck}},\
  }\href@noop {} {\bibfield  {journal} {\bibinfo  {journal} {Physics Today}\
  }\textbf {\bibinfo {volume} {62}},\ \bibinfo {pages} {40} (\bibinfo {year}
  {2009})}\BibitemShut {NoStop}%
\bibitem [{\citenamefont {Rycroft}(2009)}]{Rycroft2009}%
  \BibitemOpen
  \bibfield  {author} {\bibinfo {author} {\bibfnamefont {C.~H.}\ \bibnamefont
  {Rycroft}},\ }\href {\doibase 10.1063/1.3215722} {\bibfield  {journal}
  {\bibinfo  {journal} {Chaos: An Interdisciplinary Journal of Nonlinear
  Science}\ }\textbf {\bibinfo {volume} {19}},\ \bibinfo {pages} {041111}
  (\bibinfo {year} {2009})}\BibitemShut {NoStop}%
\bibitem [{\citenamefont {Paillusson}\ and\ \citenamefont
  {Frenkel}(2012)}]{paillusson2012probing}%
  \BibitemOpen
  \bibfield  {author} {\bibinfo {author} {\bibfnamefont {F.}~\bibnamefont
  {Paillusson}}\ and\ \bibinfo {author} {\bibfnamefont {D.}~\bibnamefont
  {Frenkel}},\ }\href@noop {} {\bibfield  {journal} {\bibinfo  {journal}
  {Physical {R}eview {L}etters}\ }\textbf {\bibinfo {volume} {109}},\ \bibinfo
  {pages} {208001} (\bibinfo {year} {2012})}\BibitemShut {NoStop}%
\bibitem [{\citenamefont {Robe}\ and\ \citenamefont
  {Boettcher}(2018)}]{Robe18}%
  \BibitemOpen
  \bibfield  {author} {\bibinfo {author} {\bibfnamefont {D.}~\bibnamefont
  {Robe}}\ and\ \bibinfo {author} {\bibfnamefont {S.}~\bibnamefont
  {Boettcher}},\ }\href {\doibase 10.1039/c8sm02191k} {\bibfield  {journal}
  {\bibinfo  {journal} {Soft Matter}\ }\textbf {\bibinfo {volume} {14}},\
  \bibinfo {pages} {9451} (\bibinfo {year} {2018})}\BibitemShut {NoStop}%
\bibitem [{\citenamefont {Dupuis}\ \emph {et~al.}(2005)\citenamefont {Dupuis},
  \citenamefont {Bert}, \citenamefont {Bouchaud}, \citenamefont {Hammann},
  \citenamefont {Ladieu}, \citenamefont {Parker},\ and\ \citenamefont
  {Vincent}}]{Dupuis2005}%
  \BibitemOpen
  \bibfield  {author} {\bibinfo {author} {\bibfnamefont {V.}~\bibnamefont
  {Dupuis}}, \bibinfo {author} {\bibfnamefont {F.}~\bibnamefont {Bert}},
  \bibinfo {author} {\bibfnamefont {J.~P.}\ \bibnamefont {Bouchaud}}, \bibinfo
  {author} {\bibfnamefont {J.}~\bibnamefont {Hammann}}, \bibinfo {author}
  {\bibfnamefont {F.}~\bibnamefont {Ladieu}}, \bibinfo {author} {\bibfnamefont
  {D.}~\bibnamefont {Parker}}, \ and\ \bibinfo {author} {\bibfnamefont
  {E.}~\bibnamefont {Vincent}},\ }\href {\doibase 10.1007/bf02704172}
  {\bibfield  {journal} {\bibinfo  {journal} {Pramana}\ }\textbf {\bibinfo
  {volume} {64}},\ \bibinfo {pages} {1109} (\bibinfo {year}
  {2005})}\BibitemShut {NoStop}%
\bibitem [{\citenamefont {Rodriguez}\ \emph {et~al.}(2003)\citenamefont
  {Rodriguez}, \citenamefont {Kenning},\ and\ \citenamefont
  {Orbach}}]{Rodriguez03}%
  \BibitemOpen
  \bibfield  {author} {\bibinfo {author} {\bibfnamefont {G.~F.}\ \bibnamefont
  {Rodriguez}}, \bibinfo {author} {\bibfnamefont {G.~G.}\ \bibnamefont
  {Kenning}}, \ and\ \bibinfo {author} {\bibfnamefont {R.}~\bibnamefont
  {Orbach}},\ }\href@noop {} {\bibfield  {journal} {\bibinfo  {journal} {Phys.
  Rev. Lett.}\ }\textbf {\bibinfo {volume} {91}},\ \bibinfo {pages} {037203}
  (\bibinfo {year} {2003})}\BibitemShut {NoStop}%
\bibitem [{Note1()}]{Note1}%
  \BibitemOpen
  \bibinfo {note} {In some experiments, such as for thermo-remanent relaxation
  in spin glasses~\cite {Dupuis2005,Rodriguez03}, it is necessary to actively
  perturb the system at a time $t_w$ to elicit as measurable response at a
  later time $t$. However, in simulations where one controls the state of every
  variable at $t_w$ and $t$, such perturbations are unnecessary.}\BibitemShut
  {Stop}%
\bibitem [{\citenamefont {Sibani}\ and\ \citenamefont {Dall}(2003)}]{Sibani03}%
  \BibitemOpen
  \bibfield  {author} {\bibinfo {author} {\bibfnamefont {P.}~\bibnamefont
  {Sibani}}\ and\ \bibinfo {author} {\bibfnamefont {J.}~\bibnamefont {Dall}},\
  }\href@noop {} {\bibfield  {journal} {\bibinfo  {journal} {Europhys. Lett.}\
  }\textbf {\bibinfo {volume} {64}},\ \bibinfo {pages} {8} (\bibinfo {year}
  {2003})}\BibitemShut {NoStop}%
\bibitem [{\citenamefont {Anderson}\ \emph {et~al.}(2004)\citenamefont
  {Anderson}, \citenamefont {Jensen}, \citenamefont {Oliveira},\ and\
  \citenamefont {Sibani}}]{Anderson04}%
  \BibitemOpen
  \bibfield  {author} {\bibinfo {author} {\bibfnamefont {P.}~\bibnamefont
  {Anderson}}, \bibinfo {author} {\bibfnamefont {H.~J.}\ \bibnamefont
  {Jensen}}, \bibinfo {author} {\bibfnamefont {L.~P.}\ \bibnamefont
  {Oliveira}}, \ and\ \bibinfo {author} {\bibfnamefont {P.}~\bibnamefont
  {Sibani}},\ }\href@noop {} {\bibfield  {journal} {\bibinfo  {journal}
  {Complexity}\ }\textbf {\bibinfo {volume} {10}},\ \bibinfo {pages} {49}
  (\bibinfo {year} {2004})}\BibitemShut {NoStop}%
\bibitem [{\citenamefont {Sibani}\ and\ \citenamefont {Jensen}(2013)}]{SJ13}%
  \BibitemOpen
  \bibfield  {author} {\bibinfo {author} {\bibfnamefont {P.}~\bibnamefont
  {Sibani}}\ and\ \bibinfo {author} {\bibfnamefont {H.~J.}\ \bibnamefont
  {Jensen}},\ }\href
  {http://www.ebook.de/de/product/19472571/paolo_sibani_henrik_jeldtoft_jensen_stochastic_dynamics_of_complex_systems.html}
  {\emph {\bibinfo {title} {Stochastic Dynamics of Complex Systems}}}\
  (\bibinfo  {publisher} {Imperial College Press},\ \bibinfo {year}
  {2013})\BibitemShut {NoStop}%
\bibitem [{\citenamefont {Boettcher}\ \emph {et~al.}(2018)\citenamefont
  {Boettcher}, \citenamefont {Robe},\ and\ \citenamefont
  {Sibani}}]{Boettcher18b}%
  \BibitemOpen
  \bibfield  {author} {\bibinfo {author} {\bibfnamefont {S.}~\bibnamefont
  {Boettcher}}, \bibinfo {author} {\bibfnamefont {D.~M.}\ \bibnamefont {Robe}},
  \ and\ \bibinfo {author} {\bibfnamefont {P.}~\bibnamefont {Sibani}},\ }\href
  {\doibase 10.1103/physreve.98.020602} {\bibfield  {journal} {\bibinfo
  {journal} {Physical Review E}\ }\textbf {\bibinfo {volume} {98}},\ \bibinfo
  {pages} {020602} (\bibinfo {year} {2018})}\BibitemShut {NoStop}%
\bibitem [{\citenamefont {Sibani}\ and\ \citenamefont
  {Boettcher}(2018)}]{Sibani18}%
  \BibitemOpen
  \bibfield  {author} {\bibinfo {author} {\bibfnamefont {P.}~\bibnamefont
  {Sibani}}\ and\ \bibinfo {author} {\bibfnamefont {S.}~\bibnamefont
  {Boettcher}},\ }\href {\doibase 10.1103/physrevb.98.054202} {\bibfield
  {journal} {\bibinfo  {journal} {Physical Review B}\ }\textbf {\bibinfo
  {volume} {98}},\ \bibinfo {pages} {054202} (\bibinfo {year}
  {2018})}\BibitemShut {NoStop}%
\bibitem [{\citenamefont {Edwards}\ and\ \citenamefont
  {Oakeshott}(1989)}]{edwards1989theory}%
  \BibitemOpen
  \bibfield  {author} {\bibinfo {author} {\bibfnamefont {S.~F.}\ \bibnamefont
  {Edwards}}\ and\ \bibinfo {author} {\bibfnamefont {R.}~\bibnamefont
  {Oakeshott}},\ }\href@noop {} {\bibfield  {journal} {\bibinfo  {journal}
  {Physica A: Statistical Mechanics and its Applications}\ }\textbf {\bibinfo
  {volume} {157}},\ \bibinfo {pages} {1080} (\bibinfo {year}
  {1989})}\BibitemShut {NoStop}%
\bibitem [{\citenamefont {Blumenfeld}\ and\ \citenamefont
  {Edwards}(2009)}]{blumenfeld2009granular}%
  \BibitemOpen
  \bibfield  {author} {\bibinfo {author} {\bibfnamefont {R.}~\bibnamefont
  {Blumenfeld}}\ and\ \bibinfo {author} {\bibfnamefont {S.~F.}\ \bibnamefont
  {Edwards}},\ }\href@noop {} {\bibfield  {journal} {\bibinfo  {journal} {The
  Journal of Physical Chemistry B}\ }\textbf {\bibinfo {volume} {113}},\
  \bibinfo {pages} {3981} (\bibinfo {year} {2009})}\BibitemShut {NoStop}%
\bibitem [{\citenamefont {Pugnaloni}\ \emph {et~al.}(2010)\citenamefont
  {Pugnaloni}, \citenamefont {S{\'a}nchez}, \citenamefont {Gago}, \citenamefont
  {Damas}, \citenamefont {Zuriguel},\ and\ \citenamefont
  {Maza}}]{pugnaloni2010towards}%
  \BibitemOpen
  \bibfield  {author} {\bibinfo {author} {\bibfnamefont {L.~A.}\ \bibnamefont
  {Pugnaloni}}, \bibinfo {author} {\bibfnamefont {I.}~\bibnamefont
  {S{\'a}nchez}}, \bibinfo {author} {\bibfnamefont {P.~A.}\ \bibnamefont
  {Gago}}, \bibinfo {author} {\bibfnamefont {J.}~\bibnamefont {Damas}},
  \bibinfo {author} {\bibfnamefont {I.}~\bibnamefont {Zuriguel}}, \ and\
  \bibinfo {author} {\bibfnamefont {D.}~\bibnamefont {Maza}},\ }\href@noop {}
  {\bibfield  {journal} {\bibinfo  {journal} {Physical Review E}\ }\textbf
  {\bibinfo {volume} {82}},\ \bibinfo {pages} {050301} (\bibinfo {year}
  {2010})}\BibitemShut {NoStop}%
\bibitem [{\citenamefont {Henkes}\ \emph {et~al.}(2007)\citenamefont {Henkes},
  \citenamefont {O'Hern},\ and\ \citenamefont
  {Chakraborty}}]{henkes2007entropy}%
  \BibitemOpen
  \bibfield  {author} {\bibinfo {author} {\bibfnamefont {S.}~\bibnamefont
  {Henkes}}, \bibinfo {author} {\bibfnamefont {C.~S.}\ \bibnamefont {O'Hern}},
  \ and\ \bibinfo {author} {\bibfnamefont {B.}~\bibnamefont {Chakraborty}},\
  }\href@noop {} {\bibfield  {journal} {\bibinfo  {journal} {Physical {R}eview
  {L}etters}\ }\textbf {\bibinfo {volume} {99}},\ \bibinfo {pages} {038002}
  (\bibinfo {year} {2007})}\BibitemShut {NoStop}%
\end{thebibliography}%

\end{document}